\documentclass{elsart}
\journal{Physics Letters B}
\usepackage{cite}
\usepackage{amssymb}
\usepackage{subeqn}
\usepackage{psfig}
\def\simgt{\lower.7ex\hbox{$\;\stackrel{\textstyle>}{\sim}\;$}}
\def\simlt{\lower.7ex\hbox{$\;\stackrel{\textstyle<}{\sim}\;$}}
\begin{document}
\begin{frontmatter}
%
\title{Correlating the Higgs and Kaon Sector CP Violations}  
\date{}
\author{M.\ Boz\ $^{a}$} and
\author{N.\ K.\ Pak\ $^{b}$}
\address{$^{a}$ Physics Department, Hacettepe University, 
06532, Ankara, Turkey}
\address{$^{b}$ Physics Department, Middle East Technical University,
06531, Ankara, Turkey} 
\begin{abstract}
Taking the most general two-doublet models with explicit CP violation in
the Higgs sector but no phase in the CKM matrix, we determine the correlation 
between the Higgs and Kaon sectors using the experimental values of $\Delta M_K$
and $\epsilon_K$. It is found that: there is a direct correlation between strength of 
FCNC couplings and the Higgs masses, and the lightest Higgs is nearly 
pure CP--even, but this small CP--odd composition gives large enough contribution
to $\epsilon_K$. 
\end{abstract}
\end{frontmatter}
\setcounter{footnote}{0} 
%
%
In the standard electroweak theory (SM), the single phase in the CKM matrix \cite{CKM}
is the unique source of CP violation which can appear only in the charged current interactions.
The physics beyond the SM, however, introduces novel sources of CP violation which 
can arise in charged as well as neutral current vertices. Indeed, CP violation 
observables offer quite useful tools \cite{biz} in search for new physics in near future $B$ 
factories such as LHC-B \cite{amol}. However, for a given model comprising the SM one
has to saturate all observables either conserving or violating the CP symmetry.

In this letter, we  assume that there is no CP violation in the SM, $i.e.$, 
CKM matrix is real. Furthermore, we take the most general two--doublet model 
with explicit CP violation. Then  we attempt to answer the following question: 
{\bf "Under such circumtances, what is the correlation between Higgs sector and 
$(B^{0},K^{0})$CP observables ?"} In fact, possibility of a real CKM matrix was discussed
in \cite{branco} by taking into account all constraints from $K^{0}$ and $B^{0}$
systems: The present experimental constraints allow for a real CKM matrix provided
that the physics beyond the SM contributes to both $K^0$ and $B_{d}^{0}$ systems.
In particular, at least $20\%$ of $\Delta M_{B_d}$ should follow from new physics
contribution modulo other conditions coming from $K^0$ system. 
Unlike \cite{branco} which assumes spontaneous CP violation,
we will restrict ourselves to explicit CP violation in the Higgs sector.
In case of spontaneous CP violation which guarantees CP invariance
of the Lagrangian before the electroweak breaking, it is possible to 
keep CKM matrix real by introducing appropriate discrete 
symmetries \cite{branco}. However, with explicit CP violation
in the Higgs sector one can not generate a real CKM matrix 
naturally. In such a case one can take reality of CKM matrix
granted and analyze $(K^{0},B^{0})$CP observables with pure
new physics CP violation coming from the Higgs sector only.
In passing, one notes that in case of spontaneous CP 
violation  electroweak baryogenesis
constraints imply a light pseudoscalar Higgs \cite{spon}. 
In summary, in what follows we will assume that CKM matrix is
pure real and all CP violation effects in the $(K^{0},B^{0})$
system spring from the explicit CP violation in the Higgs 
potential.

We adopt a general two--doublet model \cite{2doublet} with most general CP--violating
soft and hard operators:
\begin{eqnarray}
&&V(\Phi_{1},\Phi_{2})=\mu_{1}^{2}\Phi_{1}^{\dagger}\Phi_{1}
+\mu_{2}^{2}\Phi_{2}^{\dagger}\Phi_{2}
+\left[\mu_{12}^{2}\Phi_{1}^{\dagger}\Phi_{2}+h. c. \right]\nonumber\\
&&
+\lambda_{1}(\Phi_{1}^{\dagger}\Phi_{1})^{2}
+\lambda_{2}(\Phi_{2}^{\dagger}\Phi_{2})^{2}+
\lambda_{3}(\Phi_{1}^{\dagger}\Phi_{1}
\Phi_{2}^{\dagger}\Phi_{2})
+\lambda_{4}(\Phi_{1}^{\dagger}\Phi_{2}
\Phi_{2}^{\dagger}\Phi_{1})\nonumber\\
&&
+\frac{1}{2}\left[\lambda_{5}(\Phi_{1}^{\dagger}\Phi_{2})^{2}+h. c. \right]
+\left[\left(\lambda_{6}\Phi_{1}^{\dagger}\Phi_{1}+
\lambda_{7}\Phi_{2}^{\dagger}\Phi_{2}\right)\Phi_{1}^{\dagger}\Phi_{2}
+h. c. \right]
\label{higgspot}
\end{eqnarray}
where the dimensionless couplings $\lambda_{1,\cdots,4}$ are all real whereas $\lambda_{5,6,7}$
as well as the soft mass parameter $\mu_{12}^{2}$ can have nontrivial phases. One can regard 
(\ref{higgspot}) as a direct extension of the SM Higgs sector to two Higgs doublets \cite{2doublet},
or as a remnant of the supersymmetric models broken above the weak scale \cite{susy}. Altough
our framework is completely nonsupersymmetric, to reduce the unknowns in a viable manner, we
will take $\lambda_{1,\cdots,4}$ from the tree level minimal supersymmetric model  \cite{susy}
\begin{eqnarray}
\lambda_{1}=\lambda_{2}=-\frac{1}{8}(g_{1}^{2}+g_{2}^{2}), \,\,\, \lambda_{3}=-\frac{1}{4}(g_{1}^{2}-g_{2}^{2}), \,\,\
\lambda_{4}=-\frac{1}{2}g_{1}^{2},
\end{eqnarray}
but vary $\lambda_{5,6,7}$  freely. As usual we expand the neutral components of the 
Higgs doublets around the electroweak vacuum as
\begin{eqnarray}
\Phi_{i}=\frac{1}{\sqrt{2}}
\left(
\begin{array}{c}
0 \\
v_{i}+\phi_{i}+i a_{i}
\end{array}
\right)~,~~(i=1, 2) 
\label{baspil1}
\end{eqnarray}
where $v_{1}=v\ \cos \beta $, $v_2=v\ \sin \beta $ with $v=246.2~{\rm GeV}$ for correct electroweak 
breaking.

We introduce the CP--even scalars $h$, $H$, the CP--odd scalar $A$,
and the Goldstone boson $G^{0}$ (eaten by the $Z$ boson in acquiring the mass)
via the unitary rotation
\begin{eqnarray} 
\left(
\begin{array}{c} 
a_{1} \\
a_{2}\\
\phi_{1}\\ 
\phi_{2}
\end{array}  
\right)
=
\left(
\begin{array}{cccc}
\cos {\beta} & \sin {\beta} & 0 &~ 0\\
\sin {\beta} & -\cos {\beta} & 0 &~ 0\\
0 & 0& 1 &~ 0\\
0 & 0 &0 &~ 1
\end{array}
\right)\cdot
\left(
\begin{array}{c}
G_{0} \\
A\\
h\\  
H
\end{array}
\right)
\end{eqnarray}
Then, in $\left(H, h, A\right)$ basis the mass--squared matrix of the 
Higgs scalars assume the form: 
\begin{eqnarray}
\label{higgsmass}
M^{2}=\left( 
\begin{array}{ccc}
M_{11}^{2} & M_{12}^{2}&M_{13}^{2}\\
M_{12}^{2} & M_{22}^{2}&M_{23}^{2}\\
M_{13}^{2} & M_{23}^{2}&M_{a}^{2}
\end{array}\right)
\end{eqnarray}
whose entries can be expressed in terms of the parameters of potential (\ref{higgspot}) as 
\begin{eqnarray}
M_{11}&=&M_{a}^{2}-(1/2) v^{2} \sin^{2} 2 \beta \Big[\lambda_{1}+\lambda_{2}
-\left(\lambda_{3}+\lambda_{4}\right)\Big]\nonumber\\&-&2v^{2}\Big[\sin^{3}{\beta}
\Re\{\lambda_{56}\}+
\cos^{3} \beta \Re \{\lambda_{75}\}-(1/2)\sin 2 {\beta} \Re \{\lambda_{76}\}\Big]\\
M_{12}&=&v^{2}\sin 2 \beta \Big[-\lambda_{1} \cos^{2}{\beta} +\lambda_{2} \sin^{2}{ \beta}
+(1/2)\cos {2 \beta} (\lambda_3+\lambda_4) \Big]\nonumber\\
&+&v^{2}\sin {2 \beta}\Big[\cos {\beta}\Re\{\lambda_{75}\}-\sin {\beta}\Re\{\lambda_{56}\}+
\cot {2 \beta}\Re\{\lambda_{76}\}\Big]\\
M_{13}&=&v^{2}\Big[\cos {\beta}\Im\{\lambda_{75}\}-\sin {\beta}\Im\{\lambda_{56}\}\Big]\\
M_{22}&=& - 2v^{2}\Big[\lambda_{1}\cos^{4}{\beta}+
\lambda_{2}\sin^{4}{\beta}+(1/4) \sin^{2}{2 \beta} \left(\lambda_{3}+\lambda_4\right)\Big]\nonumber\\
&-& v^{2}\sin{2 \beta}\Big[\cos{\beta}\Re\{\lambda_{56}\}+ \sin {\beta}\Re\{\lambda_{75}\}+
 \Re\{\lambda_{76}\}\Big]\\
M_{23}&=&- v^{2}\Big[\cos{\beta}\Im\{\lambda_{56}\}+\sin{\beta}\Im\{\lambda_{75}\}\Big]
\end{eqnarray}
which depend on three new combinations of $\lambda_{5,6,7}$:
$\lambda_{5 6}=\sin \beta  \lambda_{5} + \cos \beta \lambda_{6}$,
$\lambda_{7 5}=\sin \beta  \lambda_{7} + \cos \beta \lambda_{5}$,
$\lambda_{7 6}=\sin^{2} \beta  \lambda_{7} + \cos^{2} \beta \lambda_{6}$.
It is understood that we have expressed the soft mass parameters
$\mu_{1}^{2}$, $\mu_{2}^{2}$ and  $\mu_{12}^{2}$ in terms of the
other parameters of the potential using the minimization condition.
In the following, the Higgs mass square matrix  (\ref{higgsmass}) will 
be diagonalized numerically via $O^{T}\cdot M^{2} \cdot O = diag.\left(M_{H_1}^{2}, M_{H_2}^{2},
M_{H_3}^{2}\right)$ where $O$ is the orthonormal Higgs mixing matrix.

Perhaps the most spectacular property of the Higgs mass--squared matrix (\ref{higgsmass}) is that the
CP--even ($H$, $h$) and CP--odd ($A$) components are mixed \cite{susy,haber}. The entries responsible for this mixing
are $M_{13}^{2}$ and $M_{23}^{2}$ both of which depend on the imaginary parts of $\lambda_{ij}$. Hence,
the explicit CP violation in the Higgs potential (\ref{higgspot}) shuffles the opposite CP components
so that the mass eigenstate Higgs scalars have no longer definite CP 
quantum numbers. In the following, 
this Higgs sector CP violation will be the unique source of CP violation in saturating the
$(K^0, B^0)$CP observables through the exchange of neutral Higgs scalars inducing flavour--changing neutral
currents (FCNC) at tree level. 

In general, FCNC's induced by the neutral Higgs exchanges depend on the model employed
for the Yukawa lagrangian \cite{fcnc}. On the other hand, in the case of spontaneous CP violation in the Higgs 
sector with nonvanishing phases of the Higgs doublets \cite{branco,branco1}, 
the requirement of a real CKM matrix leads at once to the following mass matrices for quarks
\begin{eqnarray} \Gamma^{u}=\tan\beta \left( \begin{array}{ccc} m_{u} & 0 & 0 
\\ 0 & m_{c} & 0 \\ 0 & 0 & m_{t} 
\end{array} \right) 
\label{gamu}
\end{eqnarray} 
\begin{eqnarray} 
\Gamma^{d}= \left( \begin{array}{ccc} m_{d}(\tan\beta-Sx^{2}) & 
-m_{d} S \beta_{k} & -m_{d} S \beta_{d}
\\ -m_{s} S \beta_{k} & m_{s}(\tan\beta-Sy^{2}) & -m_{s} S \beta_{s} \\ 
-m_{b} S \beta_{d} & -m_{b} S \beta_{s} & m_{b}(\tan\beta-Sz^{2}) 
\end{array} \right)
\label{gamd}
\end{eqnarray} 
where $S=\tan\beta+\cot\beta$, and $\beta_{k}=xy$, $\beta_{d}=xz$, and $\beta_{s}=yz$ with $x^2+y^2+z^2=1$
determine the strengths fo FCNC transitions in $K^0$, $B^0_d$ and $B^0_s$
systems, respectively. This flavor 
structure forbids FCNC's among up-type quarks, that is, there is no $D^0$--$\overline{D^0}$ mixing. This
particular flavor structure follows from the requirement of a real CKM 
matrix with spontaneous CP violation in the Higgs sector \cite{branco1}. 
Instead of the parametrizations for a rather general two--doublet models
\cite{fcnc}, in the following we will stick to the Yukawa structures (\ref{gamu}) and (\ref{gamd}) as an $ansatze$. In
a two--doublet model with explicit CP violation (\ref{higgspot}), such a Yukawa structure cannot be 
justified as in \cite{branco}; however, it will prove useful in analyzing $(K^{0},B^0)$CP systems.

The detailed analysis in \cite{branco} shows that for saturating the present experimental bounds, 
at least $20\%$ of $\Delta M_{B_d}$ must come from the new physics contribution. This requirement
eventually boils down to $\left(\beta_d/ \beta_K\right)^2\sim 1$, which we will assume.
Therefore, it is allowed to concentrate only on the $K^{0}$ system for the aim of saturating $\Delta M_K$,
$\epsilon_K$, and observing the resulting constraints on the Higgs sector of the theory.

The effective $\Delta S=2$ Hamiltonian for $K^0$--$\overline{K^0}$ system receives contributions 
from the exchange of charged bosons $W^{\pm}W^{\pm}$ (the SM contribution) , $W^{\pm}H^{\pm}$, $H^{\pm} H^{\pm}$ 
(two--doublet model contribution) as well as the neutral Higgs scalars $h, H, A$ (two--doublet model with tree level FCNC).
Since the CKM matrix is real, only neutral Higgs bosons contribute to $\epsilon_K$
\begin{eqnarray}
\epsilon_{K}\equiv \frac{1}{\sqrt{2}}\ \frac{\Im\ \langle \overline{K^0} | M_{12} |K^0\rangle }
{\Delta M_{K}}~,
\end{eqnarray}
whereas the neutral kaon mass difference depends on the exchange of all bosons above,
\begin{eqnarray}
\Delta {M_{K}}&\approx& 2 | \Re\ \langle \overline{K^0} | M_{12} |K^0\rangle |~.
\end{eqnarray}
The $W^{\pm}$ and $H^{\pm}$ contributions to $\langle \overline{K^0} | M_{12} |K^0\rangle$ can be found in \cite{fcnc,abbot},
and the exchange of the neutral scalars contribute by (See, for instance, the first reference in \cite{fcnc}, and \cite{branco}) 
\begin{eqnarray}
\label{m12}
\langle \overline{K^0} | M_{12} |K^0\rangle&=&\frac{G_{F}^{2}}{12 \pi^{2}}f_{K}^{2}
M_{K}\frac{m_{d}}{m_{s}}(1+\frac{m_{d}}{m_{s}})^{-1}  
\sum_{k}\left(\frac{2 \sqrt{3}\pi v M_{K}}
{M_{H_k}}\right)^{2} (U_{k,12})^{2}
\end{eqnarray}  
where $U_{k,i j}$ is defined by 
\begin{eqnarray}
U_{k,ij}=-\frac{1}{2}(S_{k,ij}-S_{k,ji}^{\ast})~,\ \ \mbox{with}\ \ 
S_{k,ij}=-\frac{{\Gamma}_{ij}^{d}}{\sqrt{m_{i}m_{j}}} (O_{1k}+iO_{3k})~
\end{eqnarray}  
using the convention that $k=1,2,3$ counts the mass eigenstate neutral Higgs scalars whereas $i,j=1,2,3$ are the generation indices. 

Equipped with the necessary formulae for the Higgs and $K^{0}$ sectors, we now turn to a numerical tracing of the
parameter space. We restrict the experimental constraints to the following bands:
\begin{eqnarray}
\label{cons}
0.98\leq |\frac{\epsilon_K}{\left(\epsilon_K\right)^{exp}}|\leq 1.02 ~~~ \mbox{and}~~~ 0.98\leq |\frac{\Delta M_K}{\left(\Delta
M_K\right)^{exp}}|\leq 1.02 
\end{eqnarray}
together with the positivity of the Higgs masses. In doing this, we wander in the parameter space varying
$|\lambda_{5}|$, $|\lambda_{6}|$, $|\lambda_{7}|$, $\mbox{Arg}[\lambda_5]$,  $\mbox{Arg}[\lambda_6]$,  $\mbox{Arg}[\lambda_7]$,
$\beta_{K}$, $\tan \beta$, and $M_a$ freely. 

We first analyze the Higgs sector within the experimental band of 
(\ref{cons}). Depicted in Fig. 1 is the scatter plot
of such an analysis for $H_1$ (top), $H_2$ (middle) and $H_3$ (bottom). Each window here shows the variation of 
the CP composition (in $\%$) of mass--eigenstate Higgs boson with its mass. It is clear that $H_1$ and $H_2$ are heavy Higgs
bosons and they have approximately alternate CP properties: $H_1$ is composed mostly of CP even elements $H$ ($"+"$) and $h$
($"\times"$) though it has also considerable CP odd composition  ($"\cdot"$) for $M_{H_1}\simlt 500~{\rm GeV}$. On the other hand,
the other heavy Higgs $H_2$ has complementary properties compared to $H_1$ with a similar range of masses. The third Higgs boson, whose
mass lies in $50 ~{\rm GeV}\simlt M_{H_3} \simlt 100~{\rm GeV}$, is the lightest of all three and it has exclusively even CP,
that is, its odd CP composition is below $0.1\%$ in the entire parameter space. As is clear from (\ref{m12}), contribution of $H_3$
to $\langle \overline{K^0} |M_{12} |K^0\rangle$ is large due to its relatively small mass; however, its CP--odd composition is
also small. 

Next, we show in Fig.2 the variation of $\beta_K$ (which is already constrained to be $< 1/2$  \cite{branco}) with the absolute 
magnitudes of $\lambda_{5, 6, 7}$ for the special case of $|\lambda_{5}|=|\lambda_{6}|=|\lambda_{7}|$. The general tendency 
of the solution points is that the required value of $\beta_K$ increases
roughly linearly with $\lambda$. Since larger the $\lambda$ larger the 
curvature of the potential, we conclude that large $\beta_K$ requires 
large Higgs masses. This 
should be the case because  $\langle \overline{K^0} |M_{12} |K^0\rangle$ is proportional to $\beta_K$ and one needs 
to suppress the FCNC due to neutral Higgs exchanges to agree with $\left(\Delta M_K\right)^{exp}$.

Depicted in Fig.3 is the dependence of $M_a$ on $\tan \beta$. As the figure suggests, higher the value of $\tan\beta$ lower the
value of $M_a$.  This, in particular, implies that $K^{0}$ constraints do not allow the Higgs sector be in the decoupling regime
in which the heavy Higgs bosons become degenerate and the lightest Higgs mass becomes maximal ($\sim 100~{\rm GeV}$), and more
importantly, the lightest Higgs approaches the pure CP-even composition \cite{susy,ben}. It is with this figure that one  arrives
at the necessity of small but sufficient CP--odd composition of the lightest Higgs for saturating $\left(\epsilon_K\right)^{exp}$.
Namely, CP--odd composition of the lightest Higgs boson,  though small, is important for $(K^0)$CP constraints due to its low mass,
enhancing $\langle \overline{K^0} |M_{12} |K^0\rangle$.

Finally, in Fig. 4 we show the dependence of ${\epsilon_K}/{\left(\epsilon_K\right)^{exp}}$ on 
${\Delta M_K}/{\left(\Delta M_K\right)^{exp}}$, when all the free parameters of the model vary. As the 
figure suggests, there are solutions in any close proximity of the experimental results.

In concluding the work, we now come back to the question mentioned in the introduction, {\bf " Yes, there is a  
correlation between the Higgs and $K^{0}$ systems"}, that is, 
\begin{enumerate}
\item The neutral Higgs exchange is sufficient to saturate $(K^0)$CP with the contribution of that Higgs boson
\begin{itemize} 
\item which is the $lightest$ of all three,
\item which is $nearly$ pure $CP$ $even$, 
\end{itemize}
\item Larger the $\Gamma^{d}_{12}$ higher the masses of the Higgs bosons to meet $\left(\Delta M_K\right)^{exp}$,
\item The Higgs sector should not slide to the decoupling limit, that is, the heavy and light Higgs bosons
should not decouple as otherwise the CP odd composition of the lightest Higgs is washed out,
\item The results  depend only on $\beta_K$ and $\beta_d$ and,
as long as  $\beta_K\sim \beta_d$ (for saturating
$\Delta M_{B_d}$) and FCNC's in the up--quark sector are neglected, one 
can take the parametrization of $\Gamma^{d}$ in (\ref{gamd})
as an ansatze for the flavour structure with free parameters
$\beta_K$ and $\beta_d$, with the constraint $|\beta_{K,s,d}|< 1/2$.
\end{enumerate}
 
{\bf Acknowledgements:} M. B. greatfully acknowledges the hospitality at the 
T\"{U}B{\.I}TAK-Feza G\"ursey Institute, Istanbul, where this work 
was initiated. We thank D. A. Demir for many invaluable discussions and suggestions.

\newpage

\begin{figure}
\centerline{\psfig{figure=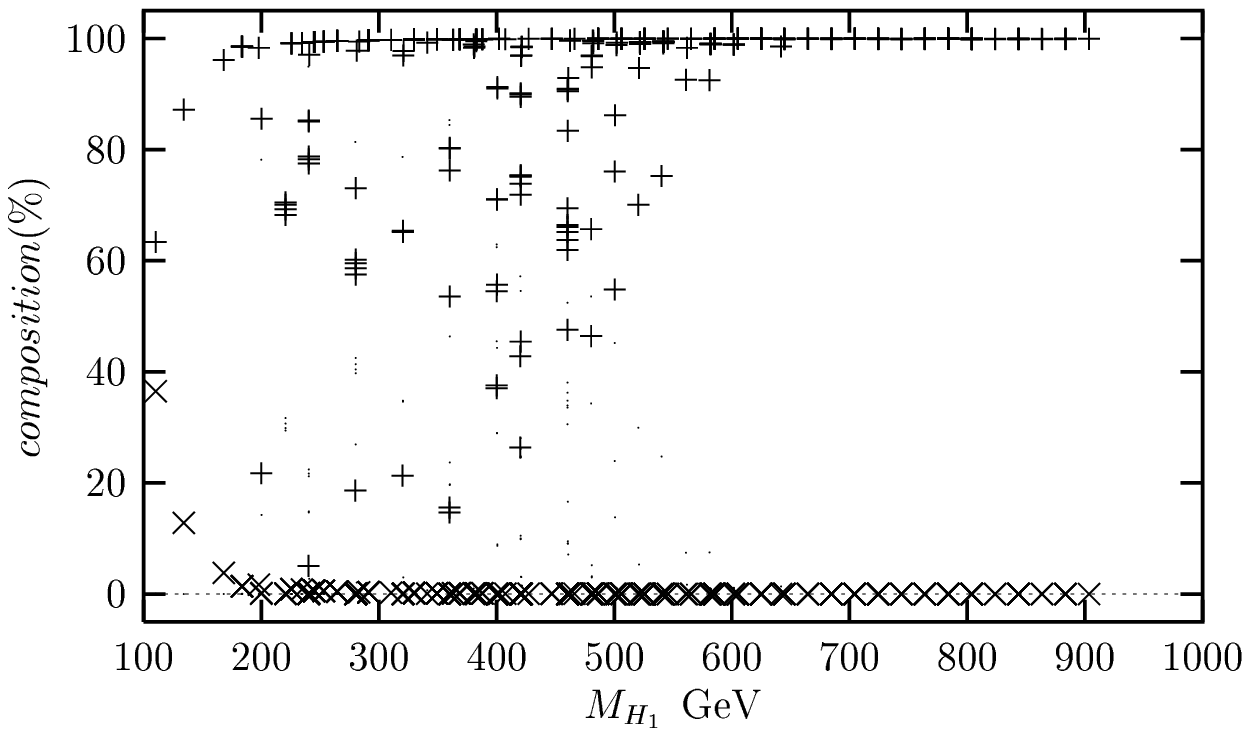}}  
\vspace{0.2cm}
\centerline{\psfig{figure=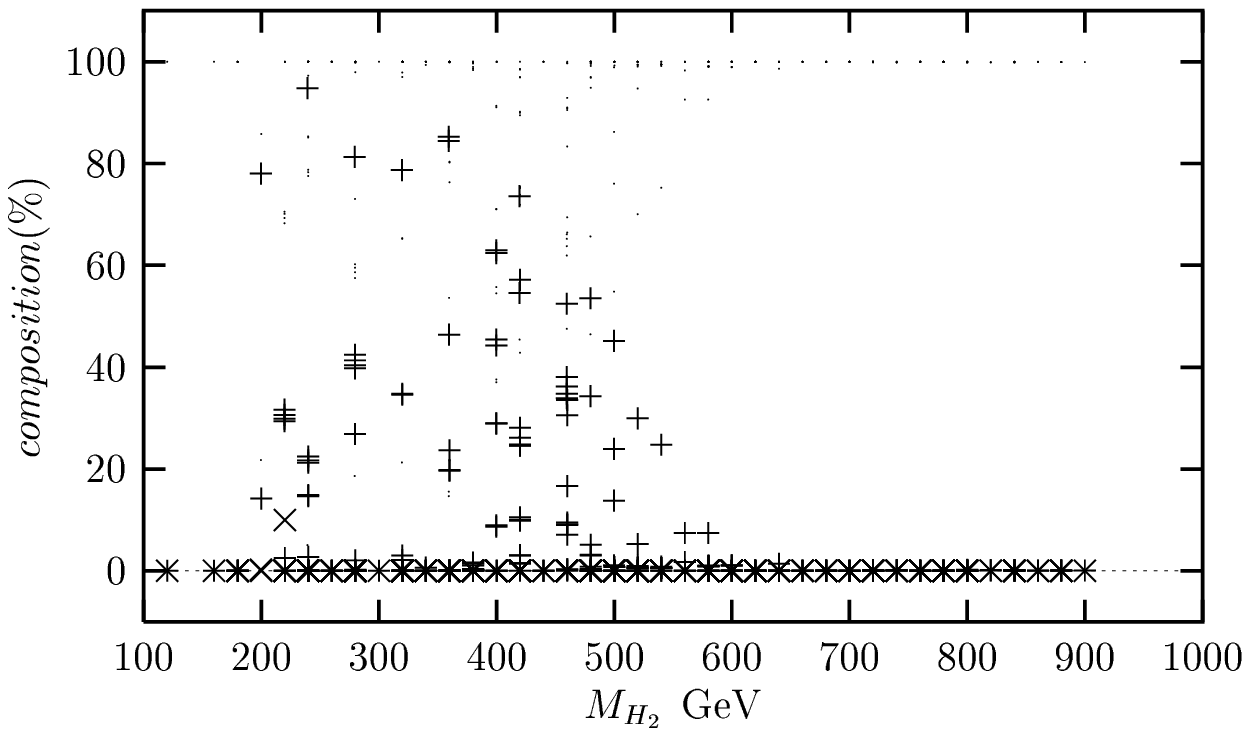}}
\vspace{0.2cm}
\centerline{\psfig{figure=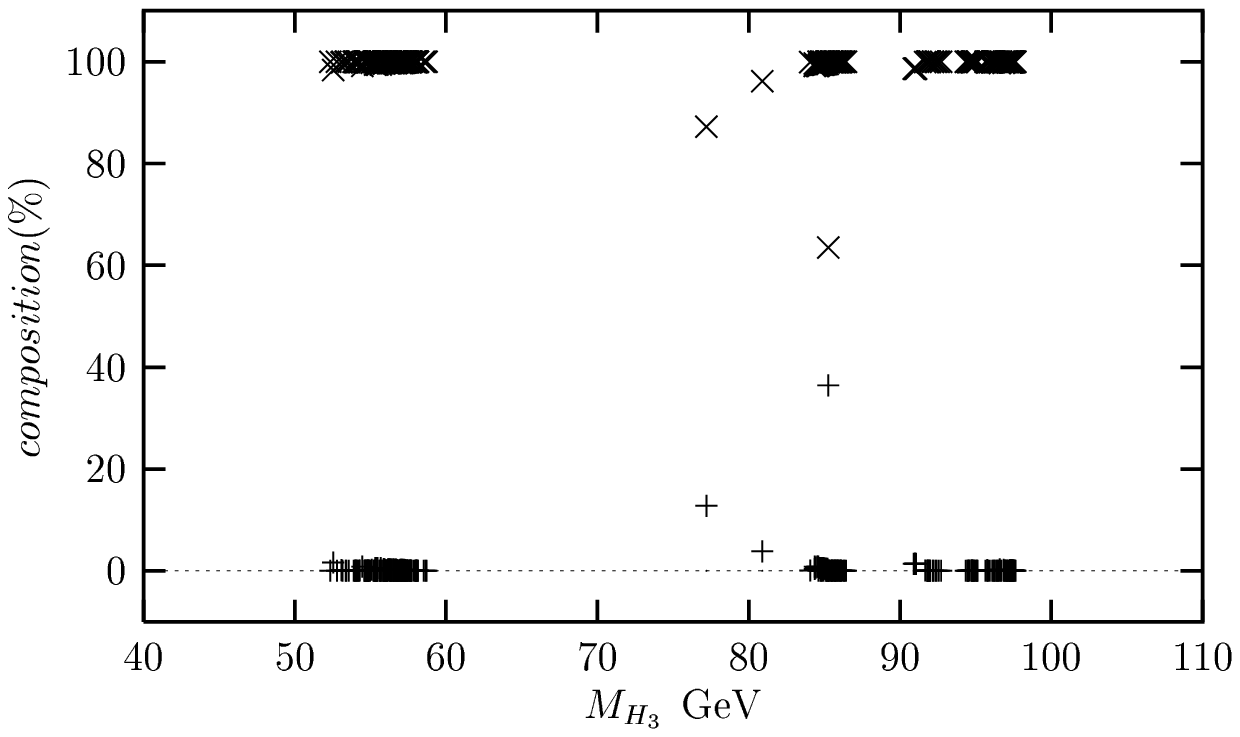}}
\caption[]{Variation of $H$ ($"+"$), $h$ ($"\times"$) and 
$A$ ($"\cdot"$) compositions ($\%$) with their masses for $H_1$ (top),
$H_2$ (middle), and $H_3$ (bottom). $H_3$ is the lightest Higgs, and its CP--odd component ($"\cdot"$)
is negligibly small.} 
\end{figure}
%
\begin{figure}
\centerline{\psfig{figure=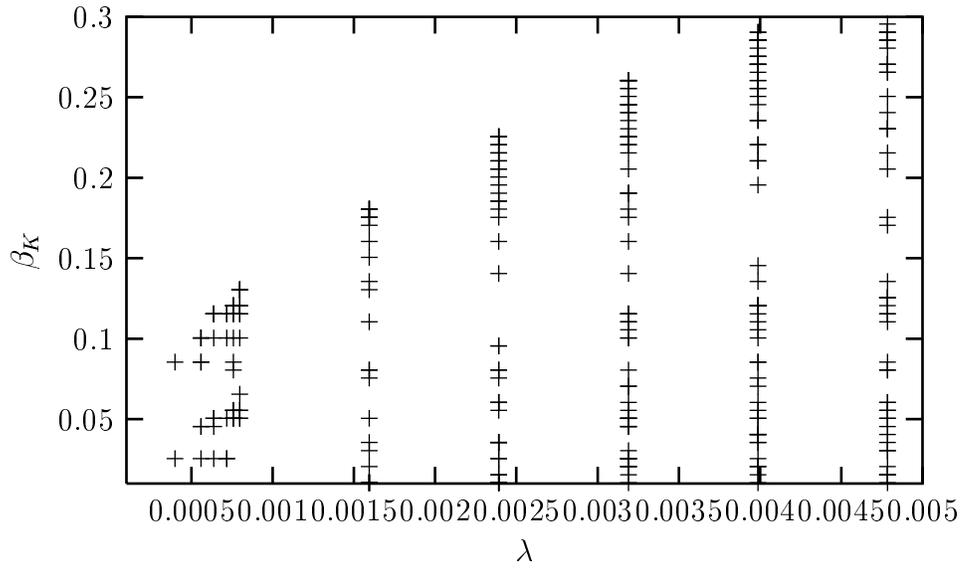}}
\caption[]{Variation of $\beta_K$ with $\lambda \equiv |\lambda_5|=|\lambda_6|=|\lambda_7|$.}
\end{figure}
%
\begin{figure}
\centerline{\psfig{figure=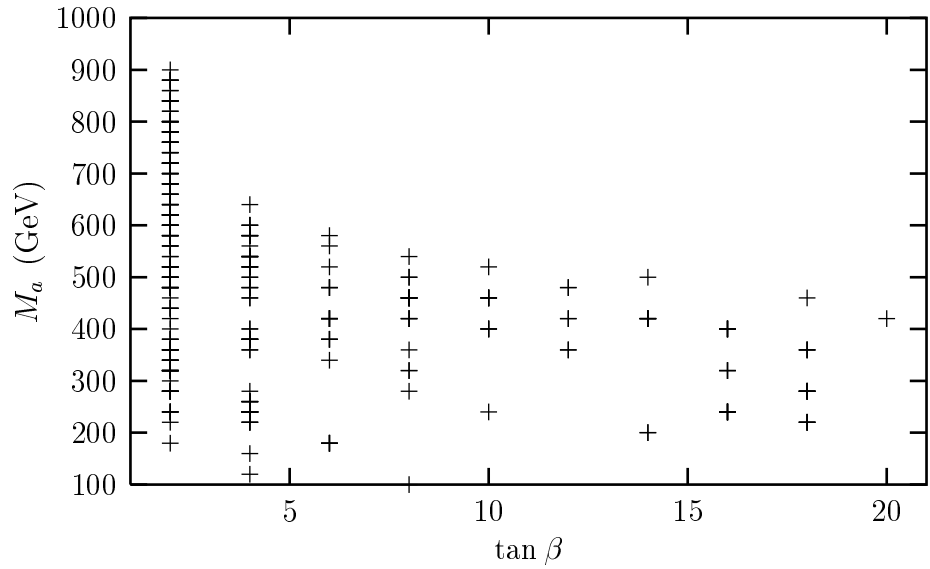}}
\caption[]{Variation of $M_a$ with $\tan \beta$}
\end{figure}

\begin{figure}
\centerline{\psfig{figure=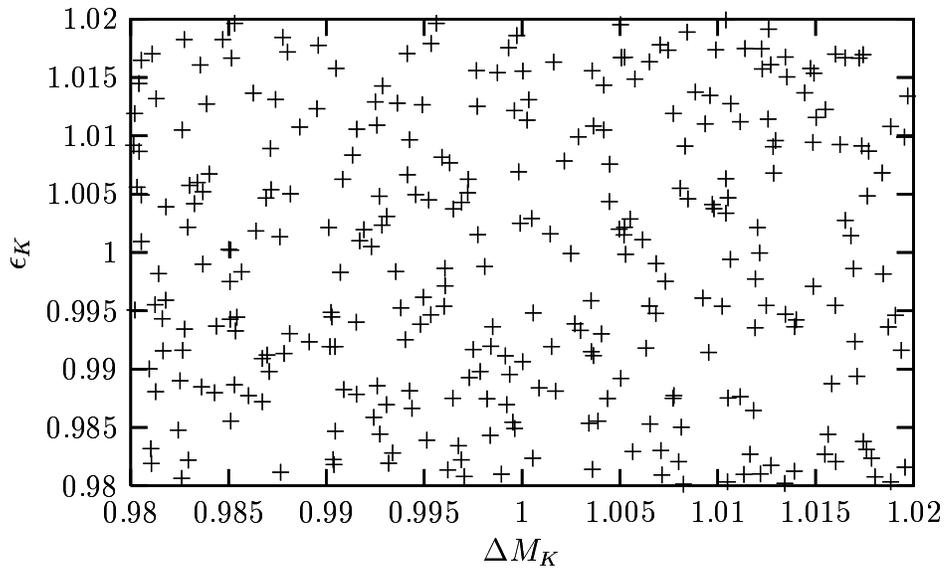}}
\caption[]{Variation of $\epsilon_K$ with $\Delta M_K$ in units of their
experimental values.}
\end{figure}

\end{document}